\documentclass[lettersize,journal]{IEEEtran}
\usepackage{amsmath,amsfonts}
\usepackage{algorithmic}
\usepackage{algorithm}
\usepackage{array}
\usepackage[caption=false,font=normalsize,labelfont=sf,textfont=sf]{subfig}
\usepackage{textcomp}
\usepackage{stfloats}
\usepackage{url}
\usepackage{verbatim}
\usepackage{graphicx}
\usepackage{cite}
\usepackage{xcolor}
\begin{document}

\title{On the Joint Optimisation of Energy Harvesting and Sensing of Piezoelectric Energy Harvesters: Case Study of a Cable-Stayed Bridge}

\author{Patricio Peralta-Braz, Mehrisadat Makki Alamdari, Elena Atroshchenko, Mahbub Hassan
\thanks{This work has been submitted to the IEEE for possible publication. Copyright may be transferred without notice, after which this version may no longer be accessible.}        
\thanks{P. Peralta-Braz, M. M. Alamdari and E. Atroshchenko are with the School of Civil and Environmental Engineering, University of New South Wales, Sydney, Australia (e-mail: e.atroshchenko@unsw.edu.au)}
\thanks{M. Hassan is with the School of Computer Science and Engineering, University of New South Wales, Sydney, Australia}
}
\markboth{ }%
{Shell \MakeLowercase{\textit{et al.}}: A Sample Article Using IEEEtran.cls for IEEE Journals}


\maketitle

\begin{abstract}
Piezoelectric Energy Harvesters (PEHs) are typically employed to provide additional source of energy for a sensing system. However, studies show that a PEH can be also used as a sensor to acquire information about the source of vibration by analysing the produced voltage signal. This opens a possibility to create Simultaneous Energy Harvesting and Sensing (SEHS) system, where a single piece of hardware, a PEH, acts as both, a harvester and as a sensor. This raises a question if it is possible to design a bi-functional PEH device with optimal harvesting and sensing performance. In this work, we propose a bi-objective PEH design optimisation framework and show that there is a trade-off between energy harvesting efficiency and sensing accuracy within a PEH design space. The proposed framework is based on an extensive vibration (strain and acceleration) dataset collected from a real-world operational cable-stayed bridge in New South Wales, Australia. The bridge acceleration data is used as an input for a PEH numerical model to simulate a voltage signal and estimate the amount of produced energy. The numerical PEH model is based on the Kirchhoff-Love plate and isogeometric analysis. For sensing, convolutional neural network AlexNet is trained to identify traffic speed labels from voltage CWT (Continuous Wavelet Transform) images. In order to improve computational efficiency of the approach, a kriging metamodel is built and genetic algorithm is used as an optimisation method. The results are presented in the form of Pareto fronts in three design spaces.

\end{abstract}

\begin{IEEEkeywords}
Piezoelectric Energy Harvester, Simultaneous Energy Harvesting and Sensing, Cable-stayed Bridge, Multi-objective Optimisation, Kriging Surrogate Model
\end{IEEEkeywords}

\section{Introduction}
\label{S:1}
While the primary function of piezoelectric energy harvesters (PEHs) is to produce electricity from various vibration sources, there is a growing interest to use the same device as a sensor to detect various environmental contexts. For example, researchers have recently demonstrated that attaching a PEH to a wearable device can not only generate electricity from human activities, but the generated electrical signal can also be directly used as a signature to characterise the activity, i.e., running, walking, jumping \cite{khalifa-tmc2018, ma2018sehs, S9246290}. Similarly, attaching the PEH device under a vibrating bridge not only generates electricity from the bridge vibrations, the generated voltage responses can be used to characterise events of interest such as detecting times of vehicle's entry and exist of the bridge~\cite{zhang2018experimental}, identifying train passages~\cite{cahill2018vibration} or even identifying bridge damage~\cite{fitzgerald2019scour}. A key benefit of using a PEH as a sensor is that it eliminates the need for a separate dedicated sensor, such as an accelerometer or strain gauge, and hence the produced energy can be utilised to more efficiently power other parts of the system, e.g. a transciever, which can be used to send data to the cloud. This concept was coined as Simultaneous Energy Harvesting and Sensing (SEHS) system in~\cite{ma2018sehs}.   

For a PEH device to act as an effective sensor, it has to be designed with the dual purpose of achieving efficient energy harvesting as well as high sensing accuracy. Although, significant progress has been made in designing efficient PEH from the energy harvesting point of view~\cite{hurtado2022shape,peralta2020parametric,lee2022piezoelectric}, little is known about the impact of PEH design parameters, such as its geometrical shape, on its sensing performance. 

In this work, we explore various PEH design spaces to understand the impact of design configurations on both energy harvesting as well as sensing accuracy. In the first study, we consider six different shapes of a cantilever-type PEH fed by an extensive vibration dataset collected from a large-scale operating cable-stayed bridge. The parametric study reveals that both energy harvesting efficiency and sensing accuracy depend significantly on the shape of the PEH, and there exist a trade-off between the harvesting and sensing performances. In the next step, we propose a joint optimisation framework and show that optimal design configurations form a Pareto front, which can be used by a designer to compromise between two objectives in each specific application.

The contributions of this paper can be summarised as follows:
\begin{enumerate}
    \item Using real field datasets from an operating cable-stayed bridge, we make the first attempt in exploring the impact of PEH geometry on both energy harvesting and sensing accuracy in the context of a vehicle speed estimation.
    \item We present a deep learning framework to evaluate the vehicle speed sensing performance of PEHs with various designs. 
    \item We report evidence, for the first time, that there are design trade-offs between energy harvesting and sensing, i.e., PEH designs that provide peak energy do not provide peak sensing accuracy. 
    \item We propose a design optimisation framework to efficiently tackle the selection of PEH geometries that address the trade-off between their energy harvesting and sensing performances. This multi-objective formulation ultimately leads to a manifold of geometries (Pareto front).
\end{enumerate}

The rest of the paper is organised as follows. The cable-stayed bridge studied in this work is presented in Section \ref{S:2}. In Section \ref{S:3}, the theoretical framework of the PEH model is explained. Next, the deep learning framework to infer the vehicle speed on a bridge using a PEH is presented in Section \ref{S:4}, and its implementation and analysis are shown in Section \ref{S:5}. Next, results are presented in Section \ref{S:5a} to evaluate the trade-off between energy harvesting and sensing performances of a PEH. Finally, a bi-objective optimisation framework is presented and implemented on an illustrative case in Section \ref{S:6} before concluding the paper in Section \ref{S:7}.

\section{Case Study: A Cable-Stayed Bridge}   
\label{S:2}

\begin{figure}[h]
\centering\includegraphics[width=0.93\linewidth]{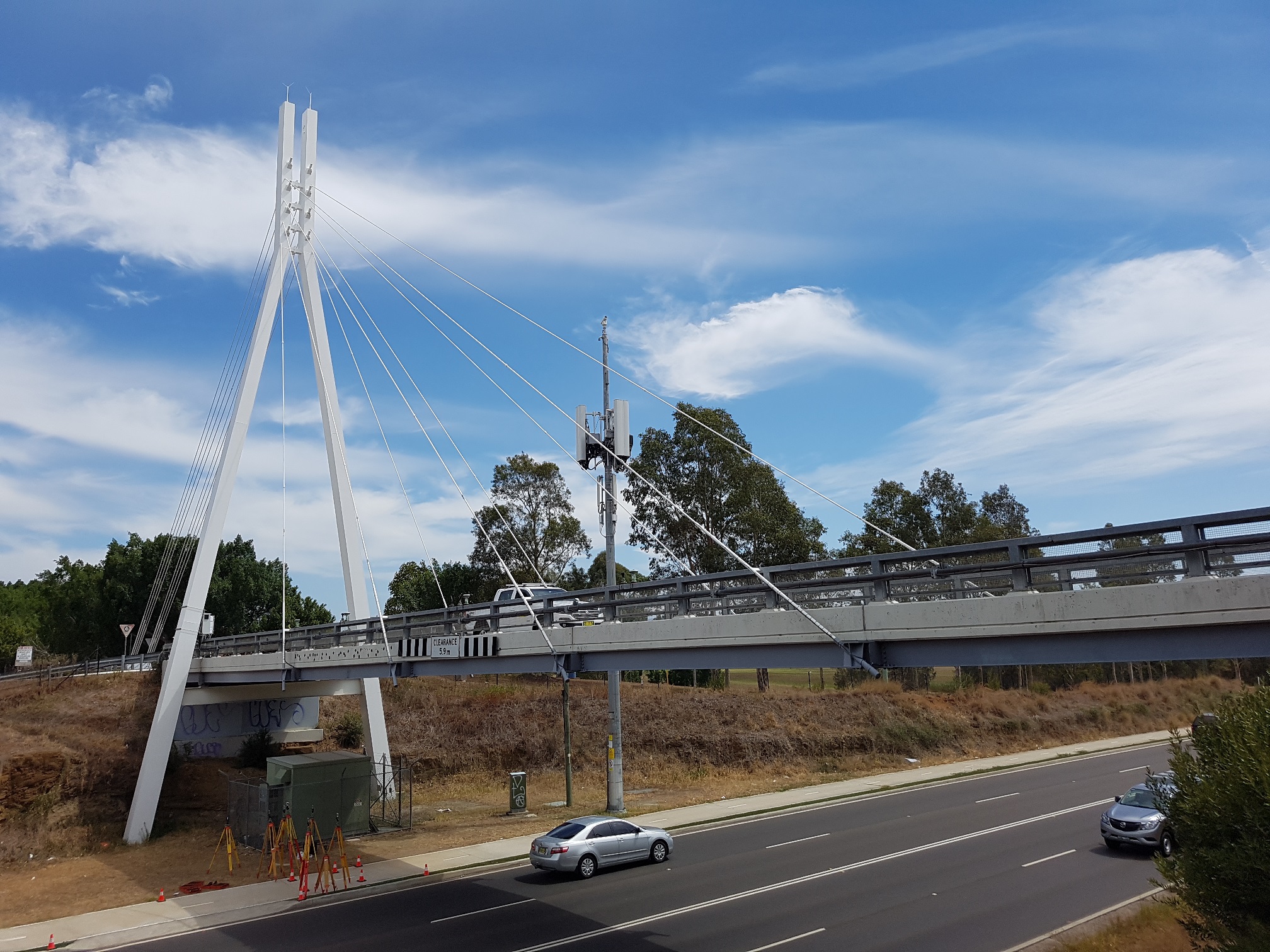}
\caption{Cable-stayed bridge located in the state of NSW, Australia \cite{kildashti2020drive}.}\label{bridge}
\end{figure}

\begin{figure}[ht]
\centering\includegraphics[width=1\linewidth]{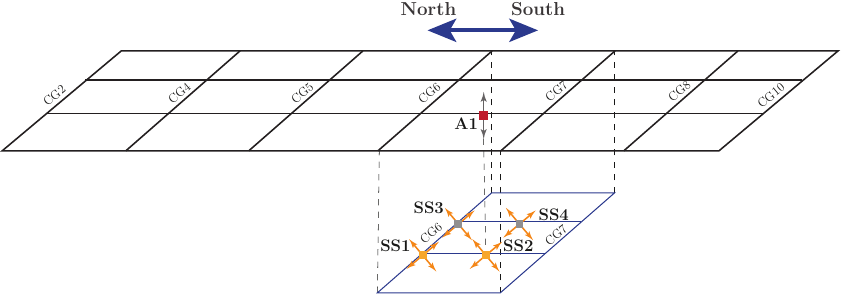}
\caption{Illustration of the sensors under the deck. The length and width of the deck are 46.22 [m] and 5.5 [m], respectively. The distance between the strain sensors is 3.975 [m]}\label{sensors}
\end{figure}

The study will be carried out on a cable-stayed bridge located over the Great Western Highway in the state of New South Wales (NSW), Australia (Fig. \ref{bridge}). The bridge has one traffic lane and one pedestrian lane with a maximum loading capacity of 30 tons. Although vehicles can travel in both directions, only traffic flow in the south-north direction is considered in this work. The bridge's deck is supported by four longitudinal girders, which are internally attached by seven cross girders (CGs), see Fig. \ref{sensors}. The bridge has an array of sensors installed under the deck, including accelerometers and strain gauges. For the purpose of this study, one accelerometer sensor (A1) and two pairs of strain gauge sensors (SS1:SS4), as shown in Fig. \ref{sensors}, are adopted. These sensors continuously measure the bridge's dynamic response at a sampling frequency of 600 Hz. The data collected from the accelerometer A1 will be employed to estimate the potential harvested energy from the passing traffic, and further infer the corresponding traffic speed. Additionally, the two pairs of strain gauge sensors are used to estimate the ground-truth velocity and consequently label samples in the training and testing database used for the Neural Network (NN) model. The vehicle speed is independently estimated from strain gauge sensors SS1-SS2 and SS3-SS4. As the sensors are installed at known distances under the bridge span, the time difference that one vehicle's axle travels from one sensor to another allows for estimating the vehicles' speed (see \cite{kalhori2017non} for more details). Although, only one pair of sensors can be used for speed estimation, in this work, we use the average between two pairs to improve the reliability of the results. 

\section{PEH IGA Model}   
\label{S:3}
\begin{figure}[h]
\centering\includegraphics[width=1\linewidth]{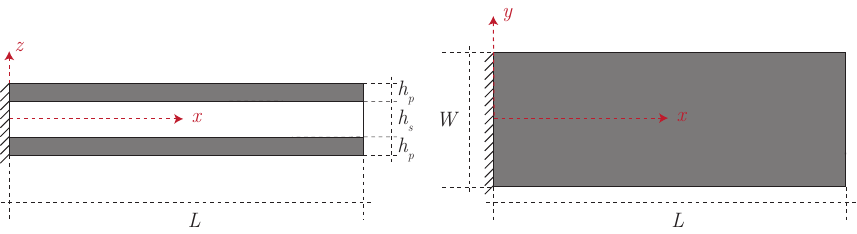}
\caption{Illustration of the piezoelectric energy harvester, modeled as a cantilever plate, consisting of two piezoelectric layers and one substructure layer.}\label{piezodevice}
\end{figure}

Several models to estimate the dynamic response of a cantilever-type piezoelectric harvester exist in the literature. In this work, we use the model based on the Kirchhoff-Love plate theory and Hamilton's principle for electro-mechanical bodies, solved numerically by the IsoGeometric Analysis (IGA). The model was proposed in \cite{peralta2020parametric} and was shown to be highly accurate at reasonable computational cost. A schematic of a PEH is shown in Fig. \ref{piezodevice}. The device is assumed to have a rectangular shape with width $W$ and length $L$, consisting of a substructure layer of thickness $h_s$ and two piezo-electric layers of thickness $h_p$ (such configuration is called bimorph), mounted on a vibrating base. To simplify all the subsequent studies, only the impact of the parameters $L$ and $W$ on energy harvesting and detection accuracy will be analysed. This means that these parameters are considered design variables, while the material parameters and thicknesses $h_s$ and $h_p$ are assumed to be constant.   

According to the IGA, B-Splines $N_I$ are used to parameterise the device’s domain and approximate the relative deflection $\textbf{w}$. We further use Modal Order Reduction approach, which consists of approximating $\textbf{w}$ by a truncated expansion of the first $K$ mode shape vectors  $\textbf{w} \approx \textbf{w}_o = \boldsymbol{\Phi}_o \boldsymbol{\eta}$. Here, $\boldsymbol{\Phi}_o\in\mathbb{R}^{N \times K}$ is the matrix which contains the $K$ first mode shape vectors $\boldsymbol{\phi}_i$ and $\boldsymbol{\eta}\in\mathbb{R}^{K\times1}$ denotes the modal coordinates. Therefore, the procedure leads to a coupled system of differential equations of the form
\begin{equation}
\label{SistEq-1R}
    \ddot{\boldsymbol{\eta}}+\textbf{c}_o\dot{\boldsymbol{\eta}}+ \textbf{k}_o\boldsymbol{\eta} - \boldsymbol{\theta}_o v(t) = \textbf{f}_o a_b(t)
\end{equation}

\begin{equation}
\label{SistEq-2R}
    C_p\dot{v}(t) + \frac{v(t)}{R_l} + \boldsymbol{\Theta}^T\boldsymbol{\Phi}_o \dot{\boldsymbol{\eta}}=0 
\end{equation}
where the equation (\ref{SistEq-1R}) corresponds to the reduced mechanical equation of motion with electrical coupling, while the equation (\ref{SistEq-2R}) corresponds to the reduced electrical circuit equation with mechanical coupling. Here, $\mathbf{k}_o\in\mathbb{R}^{K\times K}$ is the reduced stiffness matrix, $\mathbf{c}_o\in\mathbb{R}^{K\times K}$ is the reduced mechanical damping matrix, $\mathbf{f}_o\in\mathbb{R}^{K\times1}$ is the mechanical forces vector, $\boldsymbol{\Theta}\in\mathbb{R}^{N\times1}$ is the reduced electro-mechanical coupling vector, $\boldsymbol{\theta}_o\in\mathbb{R}^{K\times1}$ is the electro-mechanical coupling vector; $C_p$ is the capacitance and $R_l$ is the external electric resistance; $a_b(t)$ is the base acceleration and $v(t)$ is the output voltage; dots denote derivatives with respect to time $t$. This model was studied in detail in \cite{peralta2022design}. 

In a particular case when the base acceleration is a harmonic signal, i.e. $a_b(t) = A_b  e^{i \omega t}$, one can show that the output voltage is also harmonic, i.e. $v(t) = V_o(\omega) e^{i\omega t}$ (where $i=\sqrt{-1}$), and the Frequency Response Function (FRF), $H = H(\omega)$, can be defined to relate the amplitudes of the output voltage $V_o(\omega)$ and the excitation acceleration $A_b$ for a specific frequency $\omega$ from the equations (\ref{SistEq-1R}) and (\ref{SistEq-2R}): 
\begin{equation}
    \label{Hv_equation}
    \begin{split}
        &H = H(\omega) = \frac{V_o(\omega)}{A_b} = i\omega\left(\frac{1}{R_l}+i\omega C_p\right)^{-1} \boldsymbol{\Theta}^T\boldsymbol{\Phi}_o\times\\
         &\left(-\omega^2\textbf{I}_o+j\omega \textbf{c}_o + \textbf{k}_o + i\omega \left(\frac{1}{R_l}+i\omega C_p\right)^{-1} \boldsymbol{\theta}_o \boldsymbol{\Theta}^T\boldsymbol{\Phi}_o\right)^{-1}\textbf{f}_o
    \end{split}
\end{equation}

From the differential equations (\ref{SistEq-1R}) and (\ref{SistEq-2R}), it is also possible to estimate the voltage signal in time when the piezoelectric device is subjected to an arbitrary base acceleration $a_b(t)$, which in this study represents the acceleration measured from a vibrating bridge as a result of passing traffic. The system of differential equations can be written in its state-space form as

\begin{equation}
    \dot{\textbf{Z}} = \textbf{A}\cdot \textbf{Z} + \textbf{b} \cdot a_b(t) \label{eq:sys}
\end{equation}
where
\begin{equation*}
    \textbf{Z} = \begin{bmatrix}\boldsymbol{\eta}\\ \dot{\boldsymbol{\eta}} \\ v \end{bmatrix}, 
    \textbf{A} = \begin{bmatrix}0 & 1 & 0\\-\textbf{k}_o & -\textbf{c}_o & \boldsymbol{\theta}_o\\0 & -\frac{\boldsymbol{\Theta}^T\boldsymbol{\Phi}_o}{C_p}& -\frac{1}{C_p\cdot R_l}\end{bmatrix}, 
    \textbf{b} = \begin{bmatrix}0 \\ \textbf{f}_o \\ 0 \end{bmatrix}
\end{equation*}

Simulink ode45 solver is used to solve this system for $v(t)$. In what follows, we will investigate how the solution of system (\ref{eq:sys}) depends on the geometric parameters: length $L$ and width $W$. To emphasise that the output voltage $v(t)$ is obtained for a specific geometry $\bold{x}$, where $\bold{x} = \{L\}$ or $\bold{x} = \{L, W\}$, we use notation $v(t) = v(t, \bold{x})$. The corresponding electrical energy, $E(\bold{x})$, generated in time interval $t \in [t_1, t_2]$, can be calculated from the following equation:
\begin{equation}
    E(\bold{x}) = \int_{t_1}^{t_2}\frac{v^2(t, \bold{x})}{R_l} dt
    \label{eqEnergy}
\end{equation}

Therefore, based on the framework detailed in this section, one can obtain the voltage generated by a PEH in response to the passage of a vehicle over the bridge, characterised by the base acceleration $a_b(t)$. In the next section, a procedure to obtain the speed of a moving vehicle from the acquired voltage signal $v(t)$ is presented.

\section{Deep Learning Approach}   
\label{S:4}
In this section, a deep-learning-based framework is presented to infer a passing vehicle's speed from the voltage signal generated by a PEH installed under the bridge. In previous work, Zhou et al. \cite{zhou2021novel} presented a methodology to infer vehicle speed using the acceleration signal response of bridges purely. In the present work, this methodology is adapted and validated, incorporating the generated voltage signal from a PEH as the primary information source to infer the vehicle's speed. 

\begin{figure}[h]
\centering\includegraphics[width=1.0\linewidth]{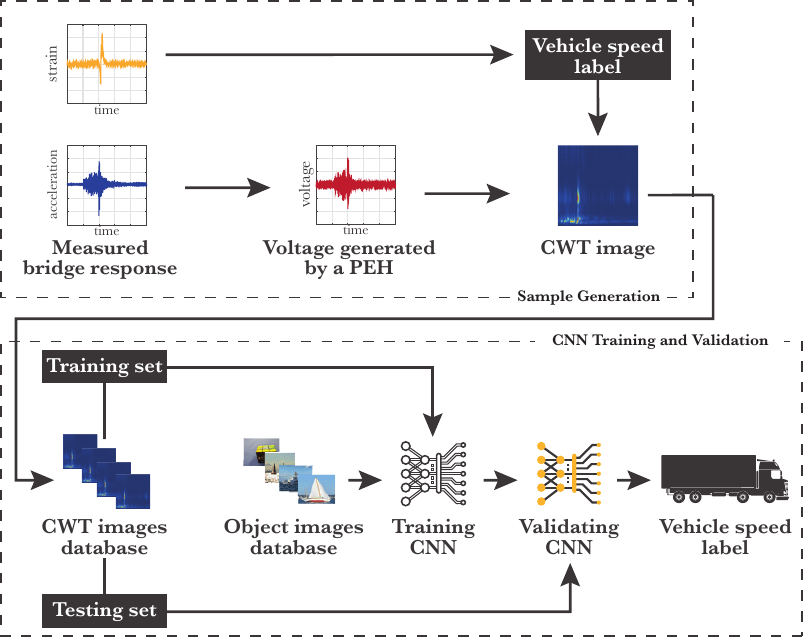}
\caption{Overview of deep learning framework comprised of two main parts: sample generation, and Convolutional Neural Network (CNN) training and testing.}\label{flow}
\end{figure}

As indicated in Fig. \ref{flow}, two main parts can be identified in this framework: sample generation and Convolutional Neural Network (CNN) training and testing. Sample generation procedure consists of the following four steps: \\
1. From the continuous bridge response database, the time window corresponding to a passing vehicle, which we call {\it an event}, is extracted. An event is defined based on the threshold criteria, i.e. the acceleration magnitude exceeding 0.15 m$/s^2$ indicates the presence of an event. The cases with multiple vehicles are not considered. Each event is taken to be 25 second. This selection is based on the span length of the bridge under investigation, and the min-max speed range on this bridge. An event is extracted in such a way, that the peak is always located at 10 seconds, i.e. the peak is identified first, then the time window starting 10 second before the peak and ending 15 second after the peak is considered. For each event, we extract the acceleration and strain signals. \\
2. The acceleration signal of an event is then used as a base acceleration input $a_b(t)$ in the PEH IGA model, described in Section \ref{S:3} and the corresponding voltage signal $v(t)$ is obtained. \\
3. For each event, the vehicle speed label is assigned from the strain measurements following the procedure explained in \cite{kalhori2017non}. \\
4. The raw time signal of the voltage $v(t)$ is post-processed using the time-frequency analysis. Continuous Wavelet Transformation (CWT) with Morlet Wavelet is employed, following the recommendation in \cite{zhou2021novel}, and the corresponding CWT image is generated.

\begin{figure}[h]
\centering\includegraphics[width=1\linewidth]{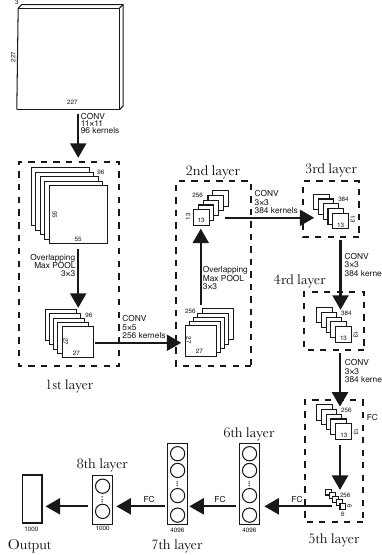}
\caption{The architecture of the adopted CNN AlexNet \cite{krizhevsky2012imagenet}.}\label{alexnet}
\end{figure}

The second part of the methodology is the Convolutional Neural Network (CNN) for training and testing. To this aim, the first step is to build a database following the procedure of sample generation. The data is divided into two sets for training and testing, respectively. The selected architecture for CNN in this paper is AlexNet \cite{krizhevsky2012imagenet} due to its excellent ability to extract local and multilevel features in different applications, and its successful implementation in vehicle classification in earlier work \cite{zhou2021novel}. AlexNet consists of eight layers, where the first five are convolutional layers, (some of them followed by max-pooling layers), and the last three are fully connected (FC) layers, as shown in Fig. \ref{alexnet}. AlexNet requires a vast database for its training; for this reason, in \cite{zhou2021novel} a Transfer Learning method \cite{pan2009survey,oquab2014learning} was proposed to reduce the size of the required databases. In this method, an alternative data is used to train lower-level layers, which are responsible for extracting the generic features of the input data. While the real data is used to train the last layers, which carry out the classification task. Finally, the trained model accuracy is estimated using the testing dataset.

\section{Sensing Accuracy}   
\label{S:5}
In this section, the methodology proposed in section \ref{S:4} is implemented and analysed. In order to use a PEH as a sensor, we expect its sensing accuracy (i.e. the accuracy of traffic speed labelling based on voltage signal $v(t)$) to be comparable with the accuracy of acceleration-based sensing (i.e. the accuracy of traffic speed labelling based on bridge acceleration signal $a_b(t)$). Hence, we first conduct a preliminary study (A), where the proposed sensing framework is applied to bridge acceleration events directly, and its accuracy is assessed. In the next study (B), we consider six configurations of PEHs, corresponding to six various values of length $L$, while all other parameters are kept constant, and label the speed of passing traffic based on the voltage signals produced by the PEHs. 

For each event, time-frequency analysis using CWT is performed to obtain two-dimensional time-frequency images, either for acceleration signal in case (A) or for voltage signal in case (B). For the sake of consistency, all images are produced considering a time limit of [5, 25] seconds and a frequency limit of [0, 200] Hz. 

In each implementation, the database is divided randomly into two sets for training and testing processes, containing 70\% and 30\% of the data, respectively. Also, the training-testing process is performed multiple times to reduce the stochastic factors. A total of 1265 samples are considered, which are classified into three different labels, as presented in Table \ref{labels}. 

\begin{table}[h]
\centering
\caption{Definition of the classification labels}
\renewcommand{\arraystretch}{1.25}
\label{labels}
\begin{tabular}{@{}llc}
\hline
Label                   &Definition                           & \# Samples\\
\hline
30 km/h                 &speed$\in$[30, 38] km/h               & 649\\
40 km/h                 &speed$\in$[42, 48] km/h               & 490\\
50 km/h                 &speed$\in$[50, 60] km/h               & 126\\
\hline
\end{tabular}
\end{table}

\subsection{Benchmark Case: acceleration-based sensing}
 
In this section, we present results for the acceleration-based sensing, which will serve as a benchmark for the voltage-based sensing. The results obtained from this study are compared with counterpart results reported in the literature performed on an experimental setup in a laboratory \cite{zhou2021novel}. 

\begin{figure}[h]
\centering\includegraphics[width=1\linewidth]{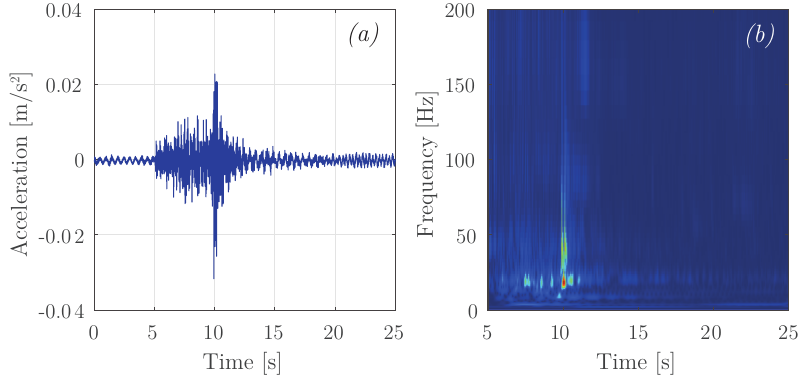}
\caption{$(a)$ An arbitrary acceleration time window induced on the bridge by a passing vehicle. $(b)$ The corresponding CWT image.}\label{acc}
\end{figure}

An example of an event and the corresponding CWT image are shown in Fig. \ref{acc}a and \ref{acc}b, respectively.  

\begin{figure}[h]
\centering\includegraphics[width=1\linewidth]{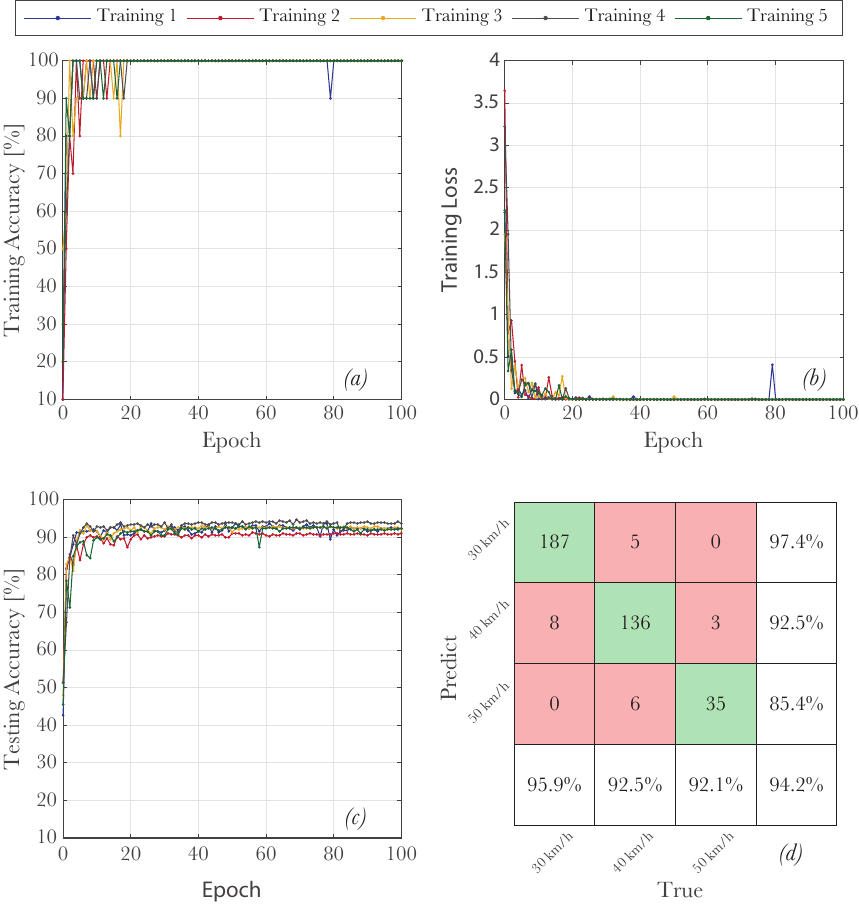}
\caption{$(a)$ Accuracy in the training process. $(b)$ Loss in the training process. $(c)$ Accuracy on the testing data across the training process. $(d)$ Confusion matrix in the testing set 5.}\label{train}
\end{figure}

The training process is carried out five times, to reduce the impact of random factors, where, each time, 70\% of the data is randomly extracted for training. In Fig. \ref{train}a and \ref{train}b, the accuracy and loss curves for five training processes are, respectively, presented. From these figures, it can be seen, that the loss continuously decreases across the epochs, and the accuracy tends to converge to 100\% at about 50 epochs. Furthermore, Fig. \ref{train}c presents the accuracy obtained from the testing dataset. The average accuracy of the five testing sets is 92.1\%, which is quite good compared to 92.9\% - 98.8\%, reported in \cite{zhou2021novel} using experimental data in contrast to field data used in the present work. Additionally, the confusion matrix of the testing set 5 is presented in Fig. \ref{train}d. 

\subsection{Study Case: voltage-based sensing}
In this section, the full framework proposed in section \ref{S:4} is employed for six different PEH designs. The materials of the devices are PZT5A and bronze \cite{peralta2020parametric}. Although materials affect the dynamic behaviour of devices and could affect detection performance, the focus of this research is to study the impact of device shape. The devices share the same geometric parameters, except for the length $L$, which is taken as 5, 10, 15, 20, 25, and 30 cm. Meanwhile, the width $W$, piezoelectric thickness $h_p$  and substructure thickness  $h_s$ are 5 cm, 0.25 mm and 0.50 mm, respectively. Other important considerations for the PEH model is that the external electrical resistance is taken as $R_l = 100 \Omega$, and the damping coefficients are assumed to be $\alpha$ = 14.65 rad/s and $\beta = 10^{-5}$ rad/s \cite{peralta2020parametric}.

\begin{figure}[h]
\centering\includegraphics[width=1\linewidth]{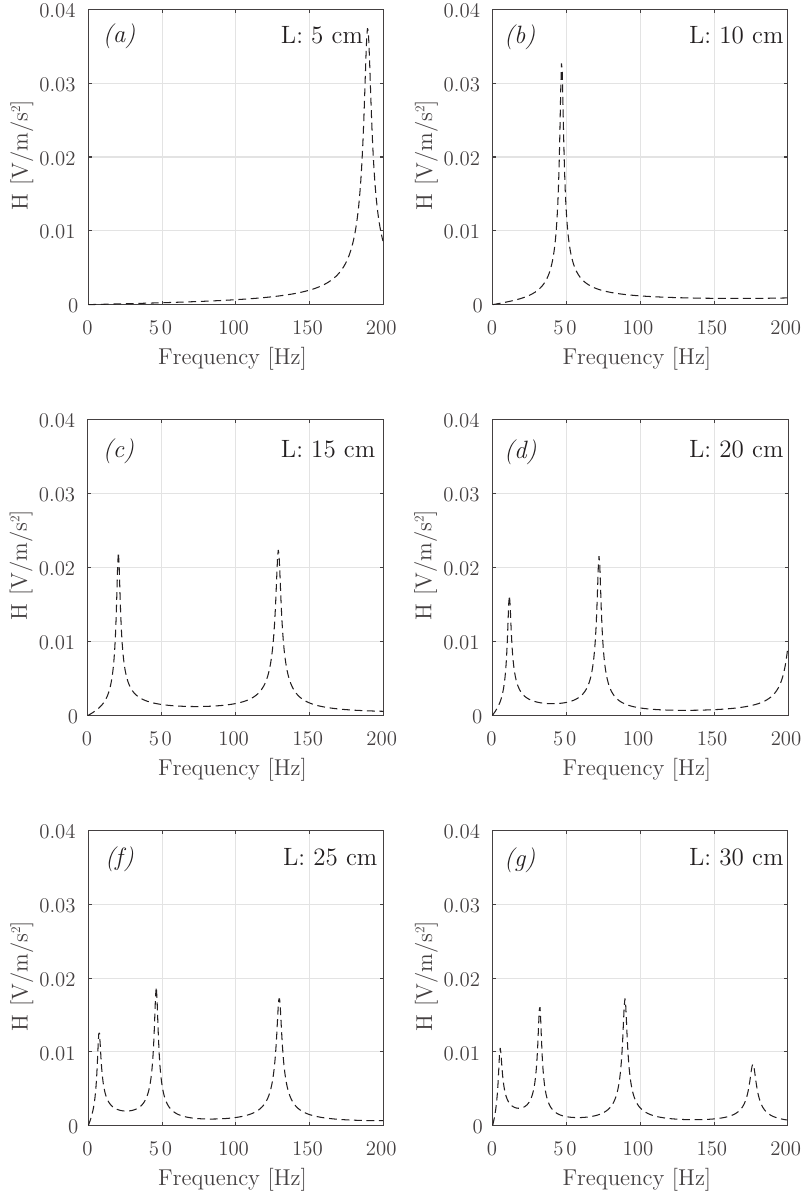}
\caption{Frequency Response Functions of the six devices with length of $(a)$ 5 cm $(b)$ 10 cm $(c)$ 15 cm $(d)$ 20 cm $(f)$ 25 cm $(g)$ 30 cm.}\label{frfs}
\end{figure}

Fig. \ref{frfs} presents the FRFs of the six studied devices, where the effect of varying the length $L$ on the dynamic response can be seen, i.e. the resonant frequencies vary as function of the length $L$. Moreover, as $L$ increases, the device tends to be more flexible with more modes appearing in the range of 0 - 200 Hz. In fact, the selection of the six PEHs for this study is based on the significant differences in their FRFs, which, as will be discussed later, result in variations in the CWT images that are directly related to sensing accuracy.

\begin{figure}[h]
\centering\includegraphics[width=0.5\linewidth]{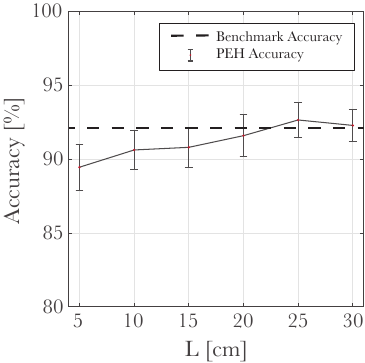}
\caption{Bar plot of accuracy from five different training processes for the six devices studied and the benchmark accuracy 92.1\% based on acceleration signal (shown in black dotted line)}\label{accu}
\end{figure}

\begin{figure}[h]
\centering\includegraphics[width=1\linewidth]{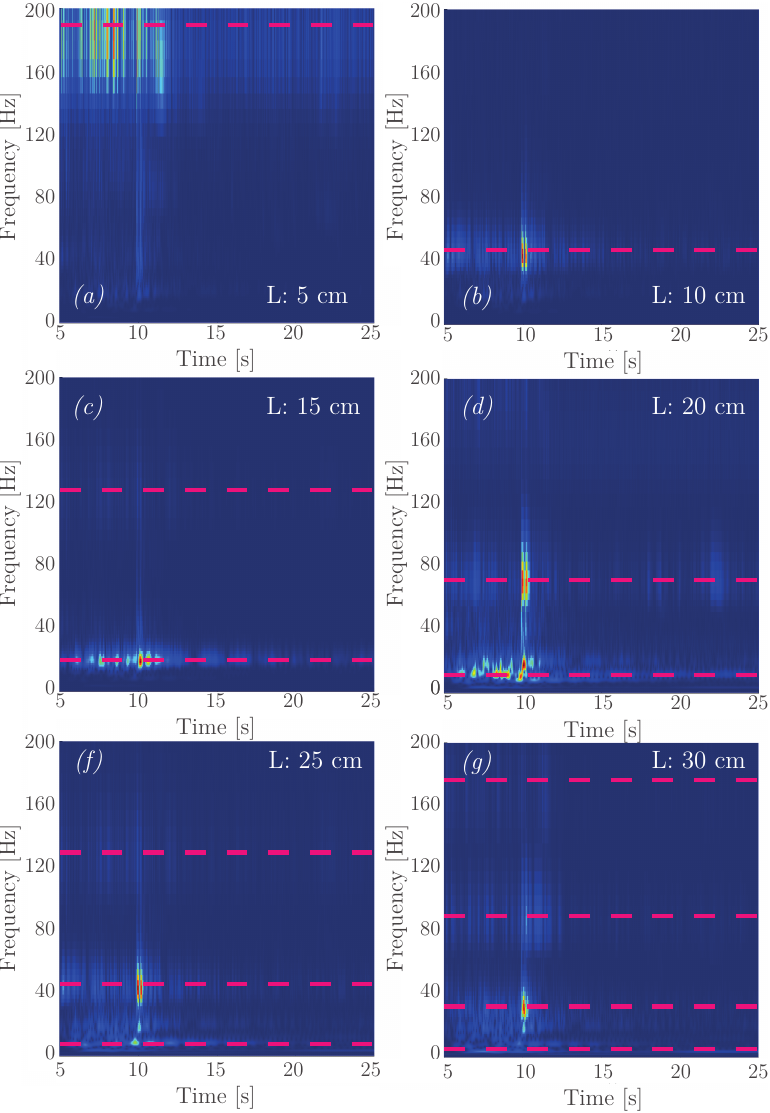}
\caption{Time-frequency image of the generated voltage (from the reference acceleration time window) obtained from the six PEH devices studied with a length of $(a)$ 5 cm $(b)$ 10 cm $(c)$ 15 cm $( d)$ 20 cm $(f)$ 25 cm $(g)$ 30 cm.}\label{CWTs}
\end{figure}

The neural network is trained fifty times for each case in order to estimate the average accuracy with greater reliability. Fig. \ref{accu} presents the mean value and error bars for the six devices considered in this study, and the benchmark accuracy, corresponding to the acceleration-based sensing, shown in the dotted black line. The PEH's accuracy varies with $L$, reaching the highest value for the device with $L = 25$ cm, which is, interestingly, higher than the benchmark accuracy. We can see that the accuracy for $L = 20$ cm and $L = 30$ cm is also comparable with the benchmark case and for $L = 5, 10, 20$ cm, the accuracy of voltage-based sensing is lower than the accuracy of acceleration-based sensing.
To interpret these results, it is necessary to consider that CNN extracts information from the time-frequency images. Fig. \ref{CWTs} presents the CWT images of the voltage signals generated from the reference acceleration window, previously presented in Fig. \ref{acc}, by the six PEHs considered in this study. The pink dashed lines represent the resonance frequencies of each device identified from the FRFs, presented in Fig. \ref{frfs}. As shown in Fig. \ref{CWTs}, the highest values of the wavelet coefficients, and consequently the most informative parts of these images, are centred close to some resonant frequencies of each PEH device, also influenced by the resonance frequencies of the bridge. Further, since devices with different length are associated with different resonance frequencies, the information they provide will differ, which consequently leads to different sensing performances. 

Next, a question arises on how the sensing accuracy and power generation are correlated. This issue will be addressed in the next section.

\section{Simultaneous Energy Harvesting and Sensing}
\label{S:5a}
In this section, the electrical energy harvested from each of the six PEH devices is studied to establish a correlation between the optimal design in terms of sensing and energy harvesting performances. Five 12-hour continuous time windows of acceleration response, between 8 am and 9 pm, are used to estimate the six candidates' average energy generated from five separate days. The energy is estimated using equation (\ref{eqEnergy}).

\begin{figure}[h]
\centering\includegraphics[width=1.0\linewidth]{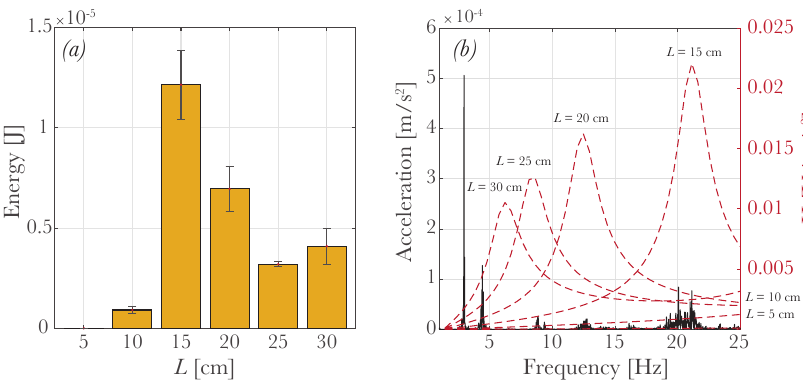}
\caption{$(a)$ Energy harvested from the six studied PEH devices obtained from five separate days. $(b)$ Acceleration frequency spectrum of one 12-hour acceleration window (shown in black), and the FRFs of the six PEH devices considered (shown in red).}\label{EnergyFig}
\end{figure}

Fig. \ref{EnergyFig}a presents the harvested energy, and the corresponding variation. From this figure, it is evident that the highest energy generation corresponds to the device with a length of 15 cm. This is explained by analysing Fig. \ref{EnergyFig}b, where the acceleration frequency spectrum of one 12-hour acceleration window from the bridge is compared with the FRFs of the six PEH devices. It can be seen that the resonance frequency of the optimal device with a length of 15 cm coincides well with dominant parts of the base excitation with frequencies around 21 Hz, which results in a higher energy generation compared to the other geometries.

\begin{figure}[h]
\centering\includegraphics[width=0.5\linewidth]{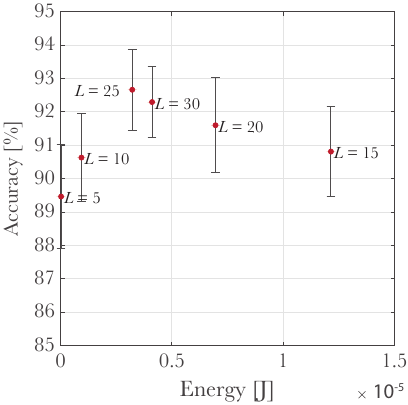}
\caption{Energy harvesting performance and the sensing accuracy of the six PEH devices.}\label{EvsS}
\end{figure}

Further, Fig. \ref{EvsS} shows the sensing performance of the six PEH devices together with the corresponding energy harvesting performance. From this figure, it is implied that there is no correlation between the optimal geometries in terms of sensing performance and energy harvesting performance, i.e. we cannot identify a device, which is optimal for both, energy harvesting and sensing. This observation opens up an opportunity for joint optimisation of sensing/harvesting performances of PEH devices.  

\section{Joint Optimisation Framework}   
\label{S:6}
The study presented in section \ref{S:5a} reveals that energy harvesting efficiency and sensing accuracy significantly depend on the shape of the PEH. Based on the results already discussed, it is possible to state that the design of a bi-functional PEH is not a trivial task because no single configuration simultaneously maximises both. In this regard, a multi-objective optimisation framework is proposed to efficiently tackle the selection of PEH geometries that address the trade-off between its energy harvesting and sensing performances. This multi-objective formulation ultimately leads to a manifold of geometries (on the boundary of the feasible objective space) called the Pareto front. When no additional information is available from the context, all the Pareto front solutions are considered equally good, despite some being better in one objective but at the same time worse in the other objective, than the other Pareto front points. The proposed optimisation framework adopts a Genetic Algorithm (GA) employing surrogate metamodels to quantify the energy harvesting performance and sensing accuracy of different geometries. Details of this optimisation scheme are given next.

\subsection{Kriging Surrogate Modeling}
One of the main challenges of the present optimisation problem is the long computational times to estimate the sensing accuracy and the energy harvesting performance of a single PEH. In order to estimate the sensing accuracy, it is necessary to build the complete training and testing database to subsequently carry out the training process on multiple occasions to estimate the expected average accuracy. While in order to quantify the energy harvesting efficiency, it is necessary to perform the numerical integration over numerous long time windows to estimate a representative expected value. When the number of design variables increases, the space of possible geometries also increases, and the optimisation problem quickly becomes computationally unfeasible. 

Surrogate modelling techniques are an attractive alternative to deal with this problem. A surrogate model (also known as a metamodel) consists of an efficient approximate relationship between the input and output of a system, using only a limited call to the original high fidelity model. Among surrogate techniques, the Kriging model \cite{zhou2020enhanced} stands out over other approaches due to its ability to indicate the level of prediction confidence at the un-sampled point, and its good performance using a reduced number of sample points \cite{qian2020sequential}. In this study, two Kriging models are implemented following the guidelines discussed in \cite{peralta2020parametric}. The input of each model corresponds to the design variables $\bold{x}$ of a PEH device. We consider three design spaces, consisting of either $L$ or $\{L, W\}$, as will be defined later. The output corresponds to one of the objective functions: the amount of produced energy, denoted by $E(\bold{x})$ or sensing accuracy, denoted by $S(\bold{x})$. Since the Kriging metamodels provide highly accurate approximations of quantities $E(\bold{x})$ and $S(\bold{x})$, in what follows we will not distinguish if the energy/sensing accuracy is obtained from the Kriging metamodel or from the original high fidelity model, and use a single notation: $E(\bold{x})$ or $S(\bold{x})$.

\subsection{Illustrative Implementation}
Three optimisation cases are performed to evaluate the effectiveness of the proposed framework and understand the trade-off between the power generation and sensing accuracy of a PEH. These cases are based on three different design spaces for the parameters $L, W$ for a bimorph PEH. The layers’ thicknesses are fixed ($h_p$ = 0.25 mm and $h_s$ = 0.50 mm). The materials of the devices are PZT5A and bronze (properties are given in Ref. \cite{peralta2020parametric}), the electrical resistance $R_l = 100 \Omega$ and the damping coefficients $\alpha$ = 14.65 rad/s and $\beta = 10^{-5}$ rad/s. The three design spaces are given by:
\begin{itemize}
    \item \textbf{Design Space 1}, $\bold{x} = \{L\}$: $L\in$  [10, 55] cm, $W$ = 5 cm
    \item \textbf{Design Space 2}, $\bold{x} = \{L\}$: $L\in$  [10, 55] cm, $W$ = $L$
    \item \textbf{Design Space 3}, $\bold{x} = \{L, W\}$: $L\in$  [10, 55] cm, $W\in$ [10, 55] cm
\end{itemize}

\begin{figure}[h]
\centering\includegraphics[width=1\linewidth]{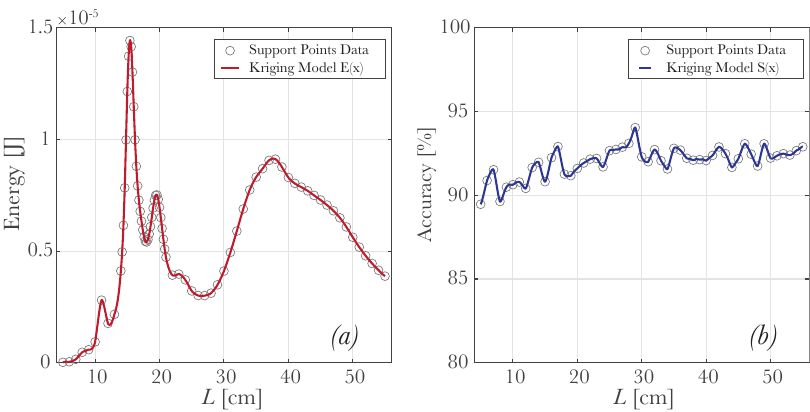}
\caption{Kriging surrogate modeling of $(a)$ harvested energy and $(b)$ sensing accuracy for Design Space 1.}\label{Kriging}
\end{figure}

Two kriging metamodels are trained for each design space to perform low computational cost output predictions for $\{E(\bold{x}),\ S(\bold{x})\}$. In the case of Design Spaces 1 and 2, 50 support points are sampled uniformly. On the other hand, as the number of design variables in Design Space 3 is two, 1000 samples are considered, which are generated with a Latin hypercube sampling algorithm. The energy harvesting performance of each sample is estimated as the average harvested energy of five 12-hour continuous time windows of acceleration response on the bridge. The sensing performance is estimated as the average accuracy obtained in fifty training-testing processes of the framework presented in Section \ref{S:4}. For illustration, Fig. \ref{Kriging} shows the data of the support points and the surrogate approximation for $E(\bold{x})$ and $S(\bold{x})$ in Design Space 1.

The optimisation problem is formalised as follows 
\begin{equation}
    \label{OptimisationEq}
    \begin{split}
         \bold{x^*}= \text{arg}\,\max\limits_{\bold{x} \in \bold{X}} \, \{ E(\bold{x}), S(\bold{x}) \}  \\ 
    \end{split}
\end{equation}

In this work, Genetical Algorithm (GA) based on a variant of NSGA-II \cite{bora2019multi,deb2011multi} is the selected optimisation method due to its good Pareto front convergence and its diversification of the candidate solutions.

\begin{figure}[h]
\centering\includegraphics[width=1\linewidth]{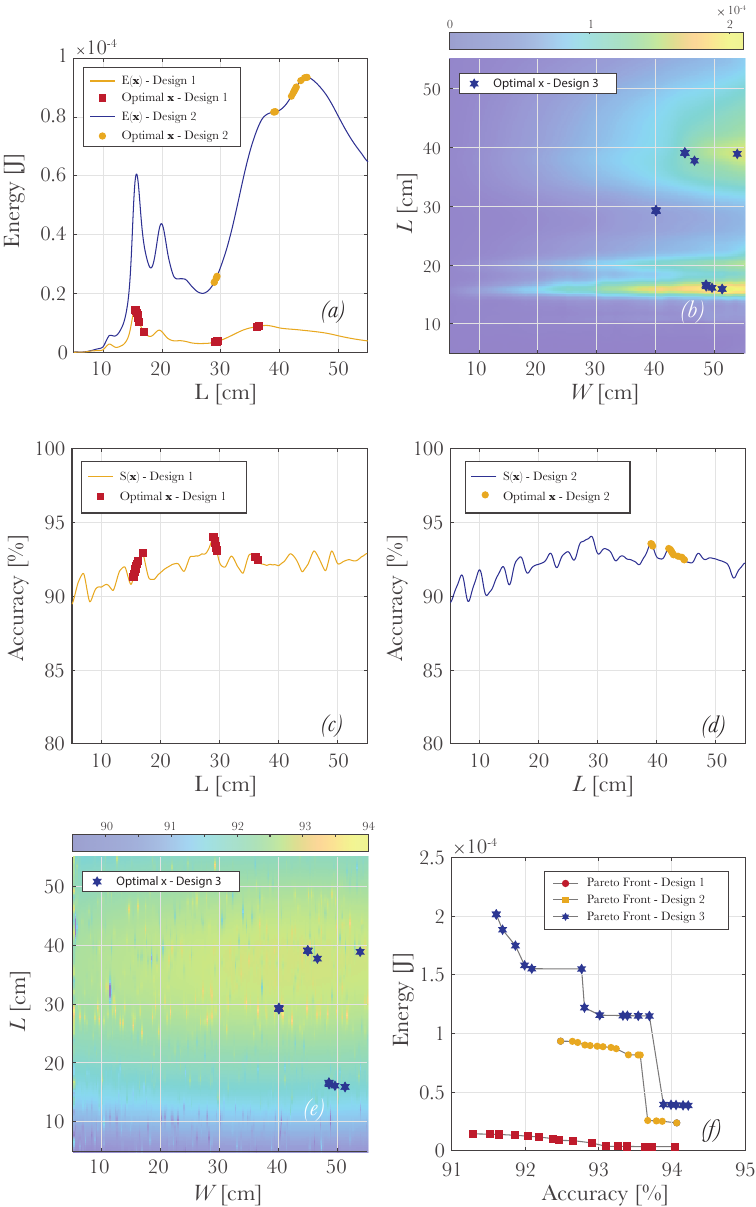}
\caption{$(a)$ Energy Harvested for Design Spaces 1 and 2. $(b)$ Energy Harvested for Design Space 3. $(c)$ Sensing Accuracy for Design Space 1. $(d)$ Sensing Accuracy for Desing Space 2. $(e)$ Sensing Accuracy for Design Space 3. $(f)$ Pareto front with optimal geometries for the maximisation of $E(\bold{x})$ and $S(\bold{x})$ in Design Spaces 1, 2 and 3.}\label{Pareto1}
\end{figure}

The optimisation results are presented in Fig. \ref{Pareto1}. Here, Fig. \ref{Pareto1}a shows $E(\bold{x})$ for Design Spaces 1 and 2, where the optimal geometries resulting from the optimisation process (Pareto front points) are also given.  Fig. \ref{Pareto1}b shows $E(\bold{x})$ for (two-dimensional) Design Space 3, together with the optimal designs from the corresponding Pareto front. 
Functions $S(\bold{x})$ for Design Spaces 1 and 2 with their optimal geometries are presented in Fig. \ref{Pareto1}c and \ref{Pareto1}d, respectively. Function $S(\bold{x})$ for Design Space 3 is presented in Fig. \ref{Pareto1}e. Finally, Fig. \ref{Pareto1}f presents the Pareto front for the two objectives $E(\bold{x})$ and $S(\bold{x})$ for the three design spaces. The competition between the objectives is evident in all three design spaces. The extremes of the Pareto front indicate the optimal designs for $E(\bold{x})$ and $S(\bold{x})$. In particular, it is possible to observe the gap between optimal values. In terms of sensing accuracy, its value varies insignificantly (within [91, 94.5]\%) across all designs. In terms of energy harvesting performance, the amount of energy varies considerably between designs, with most energy produced in Design Space 3. This is an expected results, since the two design variables in Design Space 3 are independent, so this design space is larger and contains Design Space 2. Also, the areas of devices are not comparable between Design Spaces 1, 2 and 3. It can be shown that the higher amounts of energy are produced by bigger devices in Design Space 3. 

In order to make a fair comparison between the design spaces, another optimisation process is carried out considering the energy per unit area $E(\bold{x})/A(\bold{x})$ instead of $E(\bold{x})$ as an objective, i.e. finding 
\begin{equation}
    \label{OptimisationEq}
    \begin{split}
         \bold{x^*}= \text{arg}\,\max\limits_{\bold{x} \in \bold{X}} \, \{ E(\bold{x})/A(\bold{x}), S(\bold{x}) \}  \\ 
    \end{split}
\end{equation}

\begin{figure}[h]
\centering\includegraphics[width=1\linewidth]{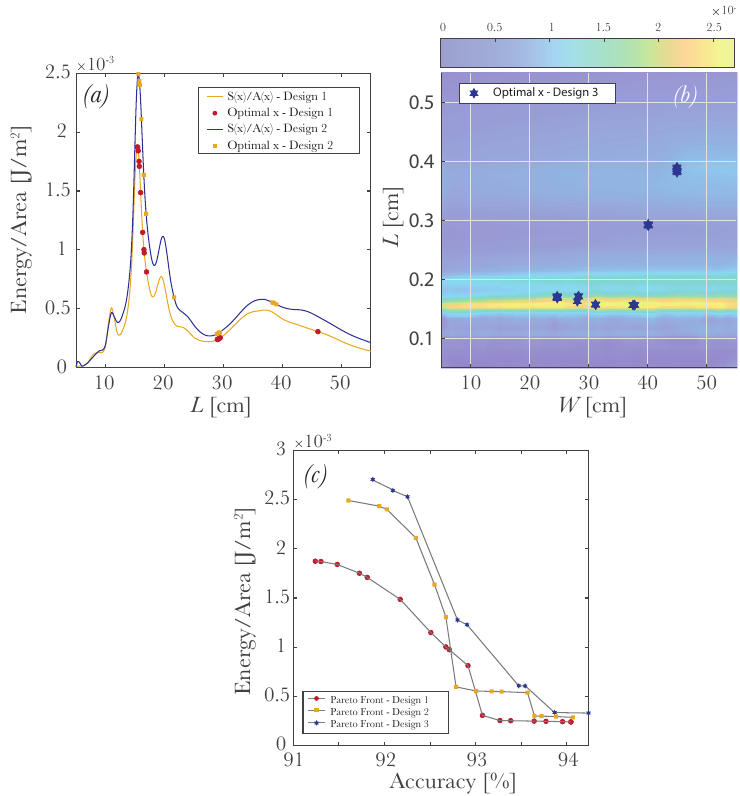}
\caption{$(a)$ Energy per unit of area for Design Spaces 1 and 2. $(b)$ Energy per unit of area for Design Space 3. $(c)$ Pareto front with optimal geometries for the maximisation of $E(\bold{x})/A(\bold{x})$ and $S(\bold{x})$ in Design Spaces 1, 2 and 3.}\label{Pareto2}
\end{figure}

Fig. \ref{Pareto2} presents the optimisation results for energy per unit area $E(\bold{x})/A(\bold{x})$ and the sensing accuracy $S(\bold{x})$. $E(\bold{x})/A(\bold{x})$ for Design Spaces 1 and 2 are shown in Fig. \ref{Pareto2}a, where the optimal geometries resulting from the optimisation process are identified. Fig. \ref{Pareto2}b shows the $E(\bold{x})/A(\bold{x})$ and optimal designs for Design Space 3. Fig. \ref{Pareto2}c presents the Pareto front for the two objectives $S(\bold{x})$ and $E(\bold{x})/A(\bold{x})$ for the three Design Spaces. The gap between both objective functions is comparable in the three designs. The best results are obtained in Design Space 3. This agrees with what was discussed earlier: Design Space 3 is larger and includes Design Space 2. This result shows that increasing the number of design variables could improve both objective functions' performance; however, the trade-off between two objectives exists in all design spaces. Finally, the variation of the optimal geometries in the Pareto front is also notorious, going from large to small devices, hence the final design choice may ultimately depend on other design requirements. 

\section{Conclusions}   
\label{S:7}

Using real vibration datasets from a cable-stayed bridge, we studied the performance of six PEH geometries in terms of both energy harvesting and sensing to correctly label the travelling speed of vehicles passing over the bridge. We found that there is no clear correlation between energy harvesting and sensing performance for a PEH. In other words, an optimum harvester may not necessarily act as an optimum sensor. 

This finding motivated the development of a rigorous design framework for joint optimisation of dual-function PEH devices that are expected to provide both energy harvesting and sensing functionality. A comprehensive study case in a real-world application was carried out to show the potential of the framework to obtain a manifold of optimal geometries (Pareto front). Ultimately, the designer is the one who decides which geometry to select based on the design requirements.

Although the optimisation framework is a good tool for understanding the trade-off between the multi-functionalities of a PEH, more studies are needed to improve the sensing performance using PEH, e.g., novel NN models. This work is the first effort to study the impact of a PEH shape on simultaneous energy harvesting and sensing and encourages future works in this direction.

\section*{Acknowledgments}
The authors wish to thank CSIRO’s Digital Productivity business unit, Data61 for providing the research data. The instrumentation and the field tests of this bridge have been planned and conducted by researchers at Data61 in collaboration with academics at University of New South Wales (UNSW). The authors also thank the Faculty of Engineering at UNSW for provision of selective seed funding under the GROW scheme. This research project was undertaken with the assistance of resources and services from the National Computational Infrastructure (NCI), which is supported by the Australian Government. The authors acknowledge support from the UNSW Resource Allocation Scheme managed by Research Technology Services at UNSW Sydney.

\bibliographystyle{IEEEtran}
\bibliography{reference.bib}

\vfill

\end{document}